\documentclass[10 pt, a4paper]{article}
\usepackage{epsfig}
\usepackage{epsf}
\usepackage{bm}                 
\usepackage{amsmath}
\usepackage{amssymb}
\usepackage{dcolumn}
\usepackage{graphicx}
\usepackage{psfrag}
\usepackage{latexsym}
\usepackage{color}

\newcommand{\beq}{\begin{equation}}
\newcommand{\eeq}{\end{equation}}
\newcommand{\beqa}{\begin{eqnarray}}
\newcommand{\eeqa}{\end{eqnarray}}

\def\Symp#1,#2,#3,#4.{\left[\left(\begin{array}{c}#1\\#2\end{array}\right),\left(\begin{array}{c}#3\\#4\end{array}\right)\right]}
\def\Vec#1,#2.{\left(\!\begin{array}{c}#1\\#2\end{array}\!\right)}
\def\vec#1,#2.{{#1\choose{#2}}}

\definecolor{redcom}{rgb}{1,0.1,0.2}
\definecolor{querycol}{rgb}{0.2,0.2,1}

\definecolor{purplerep}{rgb}{1,0.1,1}

\definecolor{green}{rgb}{0.1,0.8,1}

\begin{document}

\title{Do(es the influence of) empty waves survive in configuration space?}
\date{}
\author{}
\maketitle
\centerline{Thomas Durt\footnote{ CLARTE, Aix  Marseille  Univ,  CNRS,
Centrale  Marseille, Institut Fresnel UMR 7249,13013 Marseille, France.email: thomas.durt@centrale-marseille.fr}}


\abstract{The de Broglie-Bohm interpretation is a no-collapse interpretation, which implies that we are in principle surrounded by empty waves generated by all particles of the universe, empty waves that will never collapse. It is common to establish an analogy between these pilot-waves and 3D radio-waves, which are nearly devoided of energy but carry nevertheless information to which we may have access after an amplification process. Here we show that this analogy is limited: if we consider empty waves in configuration space, an effective collapse occurs when a detector clicks and the 3ND empty wave associated to a particle may not influence another particle (even if these two particles are identical, e.g. bosons as in the example considered here).}


\section*{Introduction.}
Several experiments were proposed in the past with the aim of revealing the existence of empty waves (see e.g. Ref. \cite{Croca} and references therein, as well as Refs. \cite{Croca90,Croca88,Crocabook,Selleri,Vigier} ). Ultimately these experiments are supposed to discriminate between with-collapse and without-collapse interpretations of quantum mechanics. Roughly speaking, in interpretations with collapse (from now on denoted CI-for collapse interpretations), the wave functions  supposedly collapses along the branch associated to the detector which clicks, at the time where the click occurs; in interpretations without collapse (from now on denoted NCI-for no collapse interpretations), all components of the wave function survive to the measurement process (detector click), even those associated to branches not associated to the detector which clicks. In many standard, textbook, presentations of the principles of the quantum theory, the collapse is associated to the so-called projection postulate which is at the core of the measurement problem. It is at the origin of many paradoxes \cite{Wheeler} such as e.g. the Schr\"odinger cat paradox and EPR paradox. Discussions and debates around these topics are still going on today, and they stimulated several proposals aimed at exploring the frontiers of the quantum realm. Some experiments aim at testing the limits of the superposition principle in the mesoscopic regime in order to reveal (the existence of) hypothetical collapse mechanisms such as spontaneous localisation processes {\it \`a la} GRW \cite{GRW}, or gravitational self-interaction {\it \`a la} Diosi-Penrose  \cite{CDWannales,CDWPRA}. If, one day, these experiments happen to lead to the detection of a non-standard signal, this would contribute to establish once for all (the validity of) with-collapse interpretations (CI). 

The quest of empty waves aims on the contrary at establishing without-collapse interpretations (NCI) such as the de Broglie-Bohm interpretation \cite{Intro,1926,bohm521,bohm522}. In this approach, particles are localised at all times, but they are guided by a wave function (pilot wave) which is not necessarily localised. The particle thus non-locally explores its surrounding via the pilot wave. Branches of the pilot wave along which the particle is not localized are called empty waves. For instance, in the double slit experiment, the particle passes through one slit, but it ``gets informed'' \cite {BH}  afterwards by the empty wave passing through the second slit about the presence of this second slit. Interference between the non-empty and empty waves influences the trajectories of the particles, which finally  gives rise to the interference pattern at the output of the two slits device. Let us now suppose that a second, identical, particle passes in the region where the empty wave associated to the first particle is located. Will its pilot wave also interfere with the empty wave? It is essentially this kind of questions that we wish to address here.

We analyse in the present paper the possibility to detect empty waves in the case where de Broglie-Bohm trajectories are described in the 3$N$ dimensional configuration space (with $N$ the number of particles). We show that previous proposals to detect empty waves implicitly privilege a description of such trajectories in the physical, 3D, space. We argue that these proposals would not make it possible to discriminate between with collapse and without collapse interpretations {\it \`a la} de Broglie-Bohm in the case where the pilot wave dynamics is defined at the level of the configuration space. However they make it possible to discriminate between, at one side, standard quantum predictions and pilot wave interpretations in configuration space and, at the other side, pilot wave interpretations in physical, 3D, space.

The paper is structured as follows.

-In section \ref{sec1} we describe an experimental scheme  described in Ref. \cite{Croca,Crocabook} in order to reveal the existence of empty waves which involves pairs of photons identically prepared in a parametric down conversion process. 

-Then, in section  \ref{sec2}, we introduce a two-photon wave function \cite{SM} in order to provide a satisfactory, fully quantum, description of the device outlined in section \ref{sec1}. In particular, the de Broglie-Bohm trajectories of the photons are seen, in this approach, to evolve in configuration space and not in the usual, 3D, physical space. Two photon interferences in configuration space are shown in the same section to play an essential role for understanding Hong Ou Mandel's (HOM) experiment \cite{HOM}.

-We show in section  \ref{sec3} that the influence of empty waves vanishes provided we describe the pilot wave dynamics in configuration space.

-In section  \ref{sec4} we distinguish a fine structure in the de Broglie-Bohm interpretation \cite{Intro}: Bohm was, according to us, the first to recognize the prior role of entanglement regarding non-local correlations between trajectories in configuration space as well as the role of decoherence during the measurement process. Bohm was in particular fully aware of the necessity to develop the pilot wave interpretation in configuration space and never questioned this peculiarity;  de Broglie on the contrary has never been very enthusiastic about the idea of a dynamics in configuration space, and persistently tried to develop a pilot wave in real, 3D space, in the spirit of the semi-classical theory outlined here in section \ref{sec1}. Several clues indicate that, actually, de Broglie was always reluctant to recognize the long range implications of entanglement \cite{dBjalon,bell}, although he was the first in 1926 to recognize these implications at the level of the guidance equation \cite{1926}. All this  brings us to discriminate 3D NCI {\it \`a la} de Broglie and 3$N$D NCI {\it \`a la} Bohm (where 3$N$D is the dimension of the configuration space in function of the number $N$ of particles-here we shall systematically consider pairs of photons so that $N$=2, and 3$N$D=6).

-The last section (\ref{sec5}) is devoted to conclusions. 

-Last but not least, we propose in appendix 1 a simplified experimental set-up, very close in mind to the HOM-dip experiment  \cite{HOM} aimed at revealing at lower cost the validity of the 3D NCI {\it \`a la} de Broglie outlined in section \ref{sec1} in the case where some phase coherence would be present between the two identical photons. We finally show in appendix 2 that if we use a laser source, that is to say, coherent states instead of pairs of equivalent photons, the predictions are the same that we use orthodox quantum optics (CI), NCI {\it \`a la} Bohm,  NCI {\it \`a la} de Broglie, or even Maxwell theory.
\section{Generation and detection of empty waves: a previous proposal.\label{sec1}}

\begin{figure}[hhh!]
  \centering
  \includegraphics[scale=0.5]{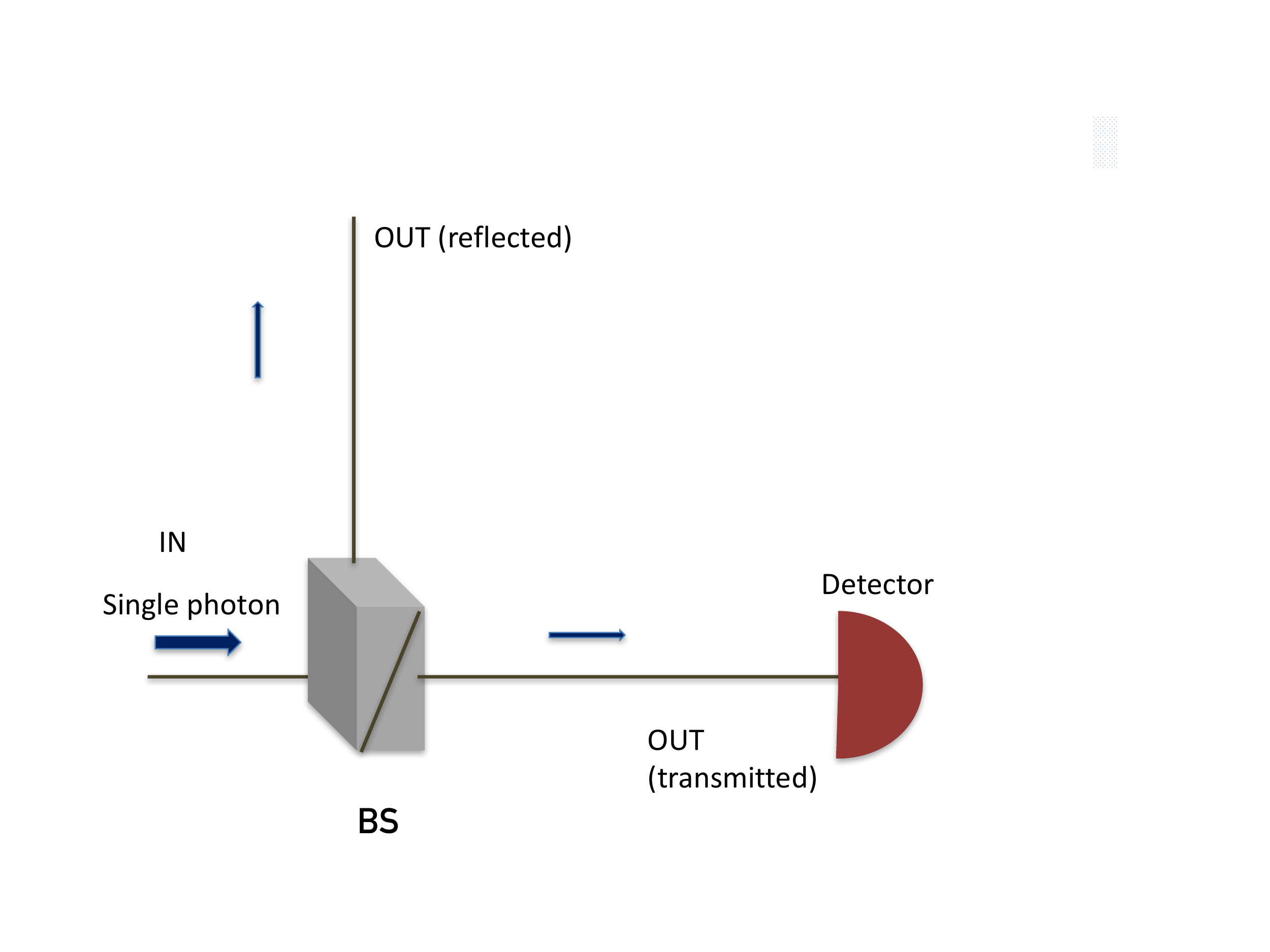}
  \caption{Generator of empty waves.}
  \label{fig1}
\end{figure}

To generate empty waves it suffices to detect a particle along a given branch of the wave function. Empty waves will de facto be created along other branches, in the case of NCI. In the case of CI, these empty waves will disappear once the particle gets detected, in virtue of the projection postulate. A simple set up shown in figure \ref{fig1} consists for instance of sending a single photon in a beamsplitter, with one detector at the output of, say, the transmitted channel. When the detector clicks, an empty wave is then created at the output of the reflected channel.
 \begin{figure}[hhh!]
  \centering
  \includegraphics[scale=0.35]{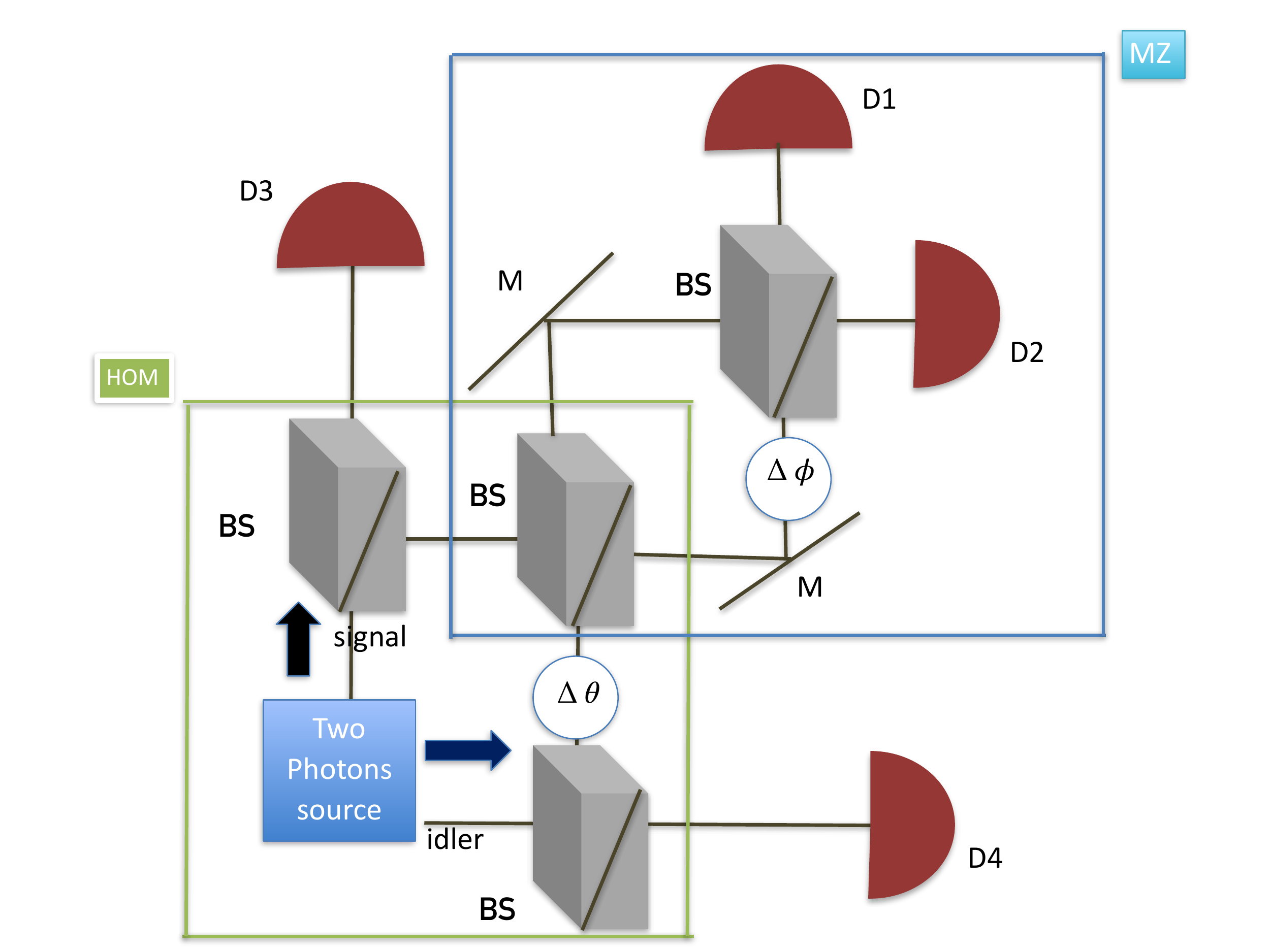}
  \caption{Generation and detection of empty waves.}
  \label{fig2a}
\end{figure}

Once an empty wave is generated, the next step is to reveal its existence.  

\begin{figure}[hhh!]
  \centering
  \includegraphics[scale=0.25]{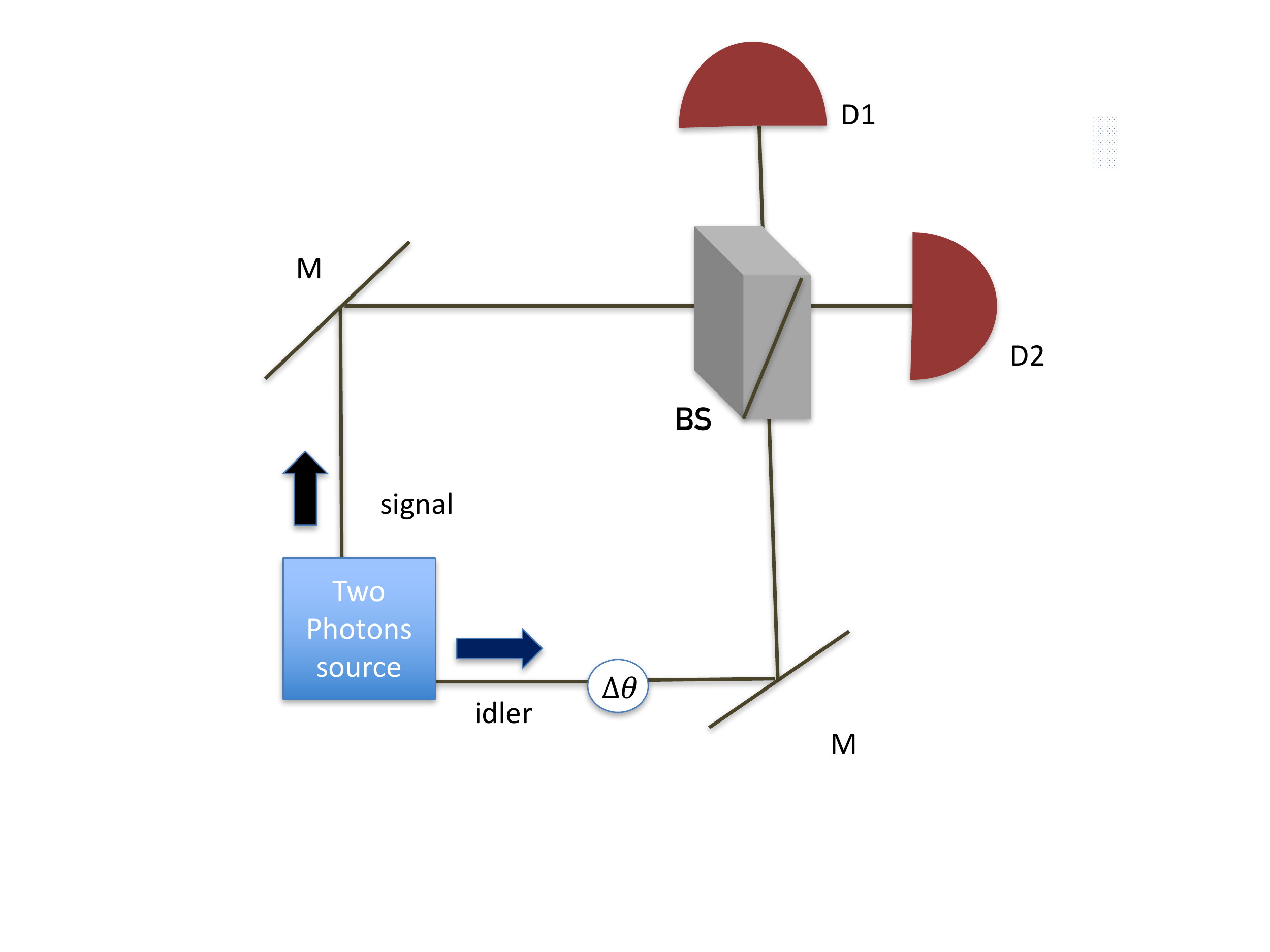}
  \caption{HOM device.}
  \label{fig2b}
\end{figure}

\begin{figure}[hhh!]
  \centering
  \includegraphics[scale=0.25]{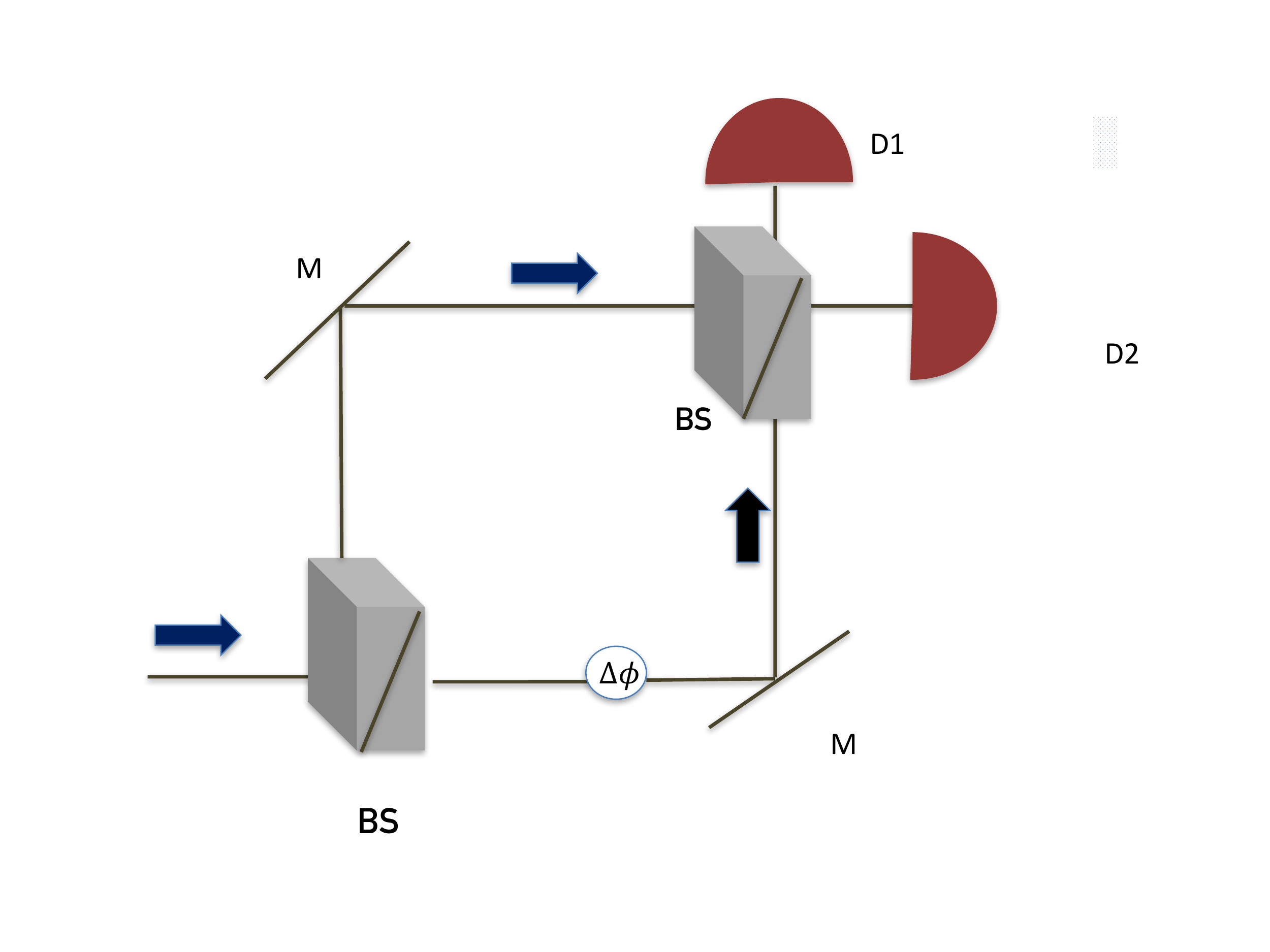}
  \caption{MZ device.}
  \label{fig2c}
\end{figure}

In figure \ref{fig2a}, we reproduce (after slight modifications) a device described in Refs. \cite{Croca,Crocabook} aimed at achieving this goal. In this device, the presence of detectors 3 and 4 aims at creating, along the scheme outlined in figure 1, an empty wave which supposedly will interfere with the other photon. Note that this scheme consists of a Hong Ou Mandel (HOM) device (see figure  \ref{fig2b} and section \ref{sec2tris}) entangled with a Mach-Zehnder (MZ) device (see figure  \ref{fig2c}). The HOM and MZ parts of the device in figure \ref{fig2a} are  respectively surrounded by a green frame and a blue frame.

For instance, if we consider the probability of firing of detectors 1 and 2, conditioned on the appearance of a click in, say, detector 4, it should be different, in case empty waves exist, from the probability of firing that would be observed when only the signal photon is sent into the Mach-Zehnder (MZ) device. 

In order to show this, let us consider what happens when only one photon enters the device, (A) in absence of (3D) empty waves and (B) in their presence. 

{\bf A) without empty waves.}

To simplify the treatment, we shall from now on assume that the dephasing $\delta \phi$ in the MZ device is equal to zero. Then, the amplitudes inside the MZ device are equal to $i\sqrt 2$ after reflection (upper arm) and $1/\sqrt 2$ after transmission (lower arm). The amplitude of the twice reflected beam (i$^2$/2) cancels the amplitude of the twice transmitted one (1/2), while the amplitudes of the once reflected-once transmitted beams are in phase and add up to unity in modulus. 

A similar result holds provided the detector 3 clicks as well as either detector 1 xor detector 2: the idler photon then enters in the MZ device vertically and leaves it vertically too. 

This means that the MZ device is transparent: the photon will leave the MZ along the arm parallel to its input arm. We expect thus coincident clicks in detectors 3 and 2, as well as in detectors 4 and 1. If no empty wave is present, as predicted in CI for instance, this is what we expect to observe. 

{\bf B) with (3D) empty waves.}

It is assumed in ref.\cite{Croca} as a preliminary work hypothesis that the signal and idler photons are incoherent in phase relatively to each other, which means that everything happens as if the dephasing $\delta \theta$ in figure  \ref{fig2a} was randomly distributed between 0 and 2$\pi$. Our approach is slightly different: we do not take for granted that the signal and idler photons are incoherent in phase relatively to each other\footnote{In appendix 1 we discuss the possibility to investigate experimentally whether phase coherence is present. Note that the experimental test proposed in appendix 1 does not make it possible to discriminate between a situation where empty waves exist and are incoherent in phase and standard predictions obtained by assuming that no empty wave is present.}. In what follows we shall thus examine two extreme cases: (B1) when maximal phase coherence is present ($\delta \theta$ is Dirac distributed) and (B2) when idler and signal photons are incoherent in phase (minimal coherence).

{\bf B1. with maximal phase coherence between idler and signal pulses.}

Contrary to the situation described in paragraph (A) above, if a click occurs in the detector 3 or 4, and that a 3-D empty wave enters the MZ device through the input opposite to the one of the surviving photon it is easy to show that, when $\delta \theta=\delta \phi=0$, the probability of firing at the output will be equal to 50 \% for each detector (1 or 2). For instance, if a single photon enters horizontally in the MZ device,  the amplitude of the upper (lower) beam inside the MZ device will both be equal to $(i+1)/2$ so that at the levels of detectors 1 and 2, the amplitudes of firing will both be equal to $(i+1)^2/2\sqrt2$ showing that the MZ device is no longer transparent in this case and actually acts as a fifty-fifty beamsplitter.

Measuring the probabilities of firing of the detectors 1 and 2, conditioned on a click in detectors 3 xor 4 constitutes thus a crucial experiment aimed at revealing the existence of empty waves of the type considered here.

One can even do better if we impose that $\delta \theta$=$\pi/2$. Then, when an empty wave is present and that a single photon enters the MZ device, the amplitude is zero along one arm inside the MZ device and the probabilities of firing of detectors 1 and 2 are the same (and equal to 1/2), disregarding the value of $\delta \phi$. The advantage of this configuration is that, when $\delta \phi$ gets varied,  the disappearance of interferences at the levels of detectors 1 and 2 reveals the existence of empty waves, while the presence of interferences reveals their non-existence\footnote{Contrary to the set up used by Mandel {\it et al.} in Ref. \cite{Mandel2} in order to reveal the existence of empty waves. }. This configuration has the advantage to be insensitive to the dephasing $\delta \phi$ between the two arms of the MZ device. It is nevertheless very sensitive to the the dephasing $\delta \theta$ at the input which is required to be exactly equal to $\pi/2$ throughout the experiment. 

Last but not least t is worth noting that, when empty waves of the type considered by Croca {\it et al.} \cite{Croca}  are present, the amplitudes of firing of detectors 1 and 2, as a function of the dephasings, can as well be computed making use of Maxwell's theory. The result is derived in appendix 2. 

{\bf B2.  with minimal phase coherence between idler and signal pulses.} 

Following Croca {\it et al.} \cite{Croca}, let us now assume that the phase difference $\delta \theta$ in figure  \ref{fig2a} is randomly distributed between 0 and 2$\pi$. Then, making use of the results of appendix 2 and averaging over $\delta \theta$ we find again that the probabilities of firing of detectors 1 and 2 are the same (and equal to 1/2), disregarding the value of $\delta \phi$. Here again, this configuration has the advantage to be insensitive to the dephasing $\delta \phi$ between the two arms of the MZ device.

 Once more, measuring the probabilities of firing of the detectors 1 and 2, conditioned on a click in detectors 3 xor 4 constitutes a crucial experiment aimed at revealing the existence of empty waves.



 \section{N-photon wave function.\label{sec2}}
At this level, we only considered 3D pilot waves, in accordance with the NCI proposed by Croca {\it et al.}; as explained in the introduction, a description of the process is possible in terms of (6D) wave functions in configuration space. Before examining this possibility, let us recall some properties of the $N$-photon wave function with a focus on the two-photon wave function.

\subsection{Spatio-temporal Wave function of a multimode single-photon.}

If we wish to describe photons propagating in space and time, a multimodal description in terms of wave packets is often more appropriate. To do so, let us introduce a one-photon state (in Schr\"odinger representation)

\begin{equation} \label{eq:OnePhoton}
\mid\psi\left(t\right)\rangle=\sum_{\lambda=\pm}\int\mathrm{d}^3k\,c_{\lambda}\left(\mathbf{k},t=0\right)\mathrm{e}^{-\mathrm{i}c\left|\left|\mathbf{k}\right|\right|t}\!\mid\!1_{\lambda,\mathbf{k}}\rangle
\end{equation}

where $\hat{a}_{\left(\lambda\right)}^\dagger\left(\mathbf{k}\right)\mid\!0\rangle=\mid\!1_{\lambda,\mathbf{k}}\rangle$ represents the quantum state of a single-mode single-photon for a (plane wave) transverse mode of wave vector $\mathbf{k}$ and polarisation $\lambda$. It is obtained after letting act the raising (creation) operator $\hat a^+_{\lambda,\mathbf{k}}$ associated to this mode on the vacuum state here denoted $\mid 0\rangle$. The vacuum state itself is the tensor-product of the vacuum states associated to all transverse modes ( $\mid\!0\rangle$=$\otimes \mid\!0_{\lambda,\mathbf{k}}\rangle, \forall \lambda,\mathbf{k}$).

The associated Glauber first order correlation  function \cite{Glauber,Scully} 
then reads 
\begin{equation} \label{eq:GeneralWaveFunction}
\begin{aligned} [b]
\bm{\psi}^E\left(\mathbf{x},t\right)=\mathrm{i}c\sum_{\lambda=\pm}\int{\mathrm{d}^3k\over (2\pi)^3} \sqrt{\left|\left|\mathbf{k}\right|\right|}\mathrm{e}^{\mathrm{i}\left(\mathbf{k}\cdot\mathbf{x}-c\left|\left|\mathbf{k}\right|\right|t\right)}c_{\lambda}\left(\mathbf{k},t=0\right)\bm{\epsilon}_{\left(\lambda\right)}\left(\mathbf{k}\right)
\end{aligned}
\end{equation}
where $\bm{\epsilon}_{\left(\lambda\right)}\left(\mathbf{k}\right)$ represents the direction of polarisation of the corresponding mode.

As is well-established in the standard theory of photodetection \cite{Scully}, the modulus squared of $\bm{\psi}^E\left(\mathbf{x},t\right)$  in the expression (\ref{eq:GeneralWaveFunction}) is proportional to the probability of detecting a photon at time $t$ in a detector located at position $\mathbf{x}$ and can thus be interpreted as a kind of single photon wave function. In analogy with the classical Poynting density, the photon wave function $\bm{\psi}^E\left(\mathbf{x},t\right)$ can also be interpreted as a single photon electric field.

\subsection{Maxwell versus Schroedinger: single photon case.\label{sec2bis}}
The quantum expression of the potential vector operator reads

$\mathbf{\hat{A}}\left(\mathbf{x},t\right)=\sum_{\lambda=\pm}\int {\mathrm{d}^3k\over (2\pi)^3} {1\over \sqrt{\left|\left|\mathbf{k}\right|\right|}}\left[\hat{a}_{\left(\lambda\right)}\left(\mathbf{k},t\right)\bm{\epsilon}_{\left(\lambda\right)}\left(\mathbf{k}\right)\mathrm{e}^{\mathrm{i}\mathbf{k}\cdot\mathbf{x}}-\hat{a}_{\left(\lambda\right)}^\dagger\left(\mathbf{k},t\right)\bm{\epsilon}_{\left(\lambda\right)}^*\left(\mathbf{k}\right)\mathrm{e}^{-\mathrm{i}\mathbf{k}\cdot\mathbf{x}}\right]$ where the cration-destruction operators $\hat{a}_{\left(\varkappa\right)}\left(\mathbf{k},t\right)=\hat{a}_{\left(\varkappa\right)}\left(\mathbf{k},t=0\right)\mathrm{e}^{-\mathrm{i}c\left|\left|\mathbf{k}\right|\right|t}$,

 and $\hat{a}^\dagger_{\left(\varkappa\right)}\left(\mathbf{k},t\right)$=$\hat{a}^\dagger_{\left(\varkappa\right)}\left(\mathbf{k},t=0\right)\mathrm{e}^{+\mathrm{i}c\left|\left|\mathbf{k}\right|\right|t}$ obey canonical commutation rules:
\begin{equation} \label{eq:aaDagger}
\left[\hat{a}_{\left(\varkappa\right)}\left(\mathbf{k},t\right),\hat{a}_{\left(\lambda\right)}^\dagger\left(\mathbf{q},t\right)\right]=\delta\left(\mathbf{k}-\mathbf{q}\right)\delta_{\varkappa\lambda}.
\end{equation} 

In analogy with Maxwell's theory, this operator is associated to the electric field operator (in the Coulomb gauge where the electric potential is equal to zero) through $\mathbf{\hat{E}}\left(\mathbf{x},t\right)={-\partial \over \partial t}\mathbf{\hat{A}}\left(\mathbf{x},t\right)$, so that

\begin{eqnarray} \label{eq:JustAMomentPos}
\mathbf{\hat{E}}\left(\mathbf{x},t\right)=\mathrm{i}c\sum_{\lambda=\pm}\int {\mathrm{d}^3k\over (2\pi)^3} \sqrt{\left|\left|\mathbf{k}\right|\right|}\left[\hat{a}_{\left(\lambda\right)}\left(\mathbf{k},t\right)\bm{\epsilon}_{\left(\lambda\right)}\left(\mathbf{k}\right)\mathrm{e}^{\mathrm{i}\mathbf{k}\cdot\mathbf{x}}-\hat{a}_{\left(\lambda\right)}^\dagger\left(\mathbf{k},t\right)\bm{\epsilon}_{\left(\lambda\right)}^*\left(\mathbf{k}\right)\mathrm{e}^{-\mathrm{i}\mathbf{k}\cdot\mathbf{x}}\right]\nonumber \\=\mathrm{i}c\sum_{\lambda=\pm}\int {\mathrm{d}^3k\over (2\pi)^3} \sqrt{\left|\left|\mathbf{k}\right|\right|}\left[\hat{a}_{\left(\lambda\right)}\left(\mathbf{k}\right)\bm{\epsilon}_{\left(\lambda\right)}\left(\mathbf{k}\right)\mathrm{e}^{\mathrm{i}(\mathbf{k}\cdot\mathbf{x}-c\vert\vert\boldsymbol{k}\vert\vert t)}-\hat{a}_{\left(\lambda\right)}^\dagger\left(\mathbf{k}\right)\bm{\epsilon}_{\left(\lambda\right)}^*\left(\mathbf{k}\right)\mathrm{e}^{-\mathrm{i}(\mathbf{k}\cdot\mathbf{x}-c\vert\vert\boldsymbol{k}\vert\vert t)}\right]
\end{eqnarray}

Similarly, the  magnetic field operator obeys $\mathbf{\hat{B}}\left(\mathbf{x},t\right)=rot \mathbf{\hat{A}}\left(\mathbf{x},t\right)$ so that

\begin{equation}
\hat{\boldsymbol{B}}(\boldsymbol{x},t)
= \mathrm{i} \sum_{\lambda=\pm}\int{\mathrm{d}^3k\over (2\pi)^3} (1/\sqrt{\left|\left|\mathbf{k}\right|\right|})\left(\boldsymbol{k}\times\boldsymbol{\epsilon}_{\left(\lambda\right)}\left(\boldsymbol{k}\right)\right)\left[\hat{a}_{\left(\lambda\right)}\left(\boldsymbol{k}\right)\mathrm{e}^{-\mathrm{i}(c\vert\vert\boldsymbol{k}\vert\vert t-\boldsymbol{k}\cdot\boldsymbol{x})}-\hat{a}_{\left(\lambda\right)}^\dagger\left(\boldsymbol{k}\right)\mathrm{e}^{\mathrm{i}(c\vert\vert\boldsymbol{k}\vert\vert t-\boldsymbol{k}\cdot\boldsymbol{x})}\right] ,\nonumber \\
\end{equation}
  It can be shown by straightforward computation that the Glauber first order correlation  function can be expressed in terms of the electric field operator through Glauber's extraction rule 
\begin{equation} \label{eq:GeneralWaveFunctionplus}
\begin{aligned} [b]
\bm{\psi}^E\left(\mathbf{x},t\right)&\equiv\langle 0\!\mid\hat{\mathbf{E}}\left(\mathbf{x},0\right)\mid\psi\left(t\right)\rangle\\
&=\sum_{\lambda=\pm}\int\mathrm{d}^3k\,\langle0\!\mid\hat{\mathbf{E}}\left(\mathbf{x},0\right)c_{\lambda}\left(\mathbf{k},t=0\right)\mathrm{e}^{-\mathrm{i}c\left|\left|\mathbf{k}\right|\right|t}\!\mid\!1_{\lambda,\mathbf{k}}\rangle\\
\end{aligned}
\end{equation}
If $N$ photons are present, the $N$ photons electric wave functions are obtained by applying $N$ times the Glauber extraction rule:
\begin{equation} \label{eq:GeneralWaveFunctionplusN}
\begin{aligned} [b]
\bm{\psi}^{E} \left(\mathbf{x_1},\mathbf{x_2},...\mathbf{x_i},...\mathbf{x_N},t\right)&\equiv {1\over \sqrt{N!}}\langle 0\!\mid\hat{\mathbf{E}}\left(\mathbf{x_1},0\right)\hat{\mathbf{E}}\left(\mathbf{x_2},0\right)...\hat{\mathbf{E}}\left(\mathbf{x_i},0\right)...\hat{\mathbf{E}}\left(\mathbf{x_N},0\right)\mid\psi\left(t\right)\rangle\end{aligned}
\end{equation}
Inspired by Maxwell theory, let us define,  making use of the magnetic field operator, the (transverse) magnetic single-photon wave function \cite{Goessens} as follows: \begin{equation}
 \bm{\psi}^B\left(\mathbf{x},t\right)  \equiv  \langle0\vert\hat{\boldsymbol{B}}^+(\boldsymbol{x},t=0)\mid \psi\left(t\right)\rangle  \quad .
   \label{eq:defEBP}
  \end{equation}
  If $N$ photons are present, the $N$ photons magnetic wave functions are obtained once again by applying $N$ times the Glauber extraction rule:
\begin{equation} \label{eq:GeneralWaveFunctionplusN}
\begin{aligned} [b]
\bm{\psi}^{B} \left(\mathbf{x_1},\mathbf{x_2},...\mathbf{x_i},...\mathbf{x_N},t\right)&\equiv {1\over \sqrt{N!}}\langle 0\!\mid\hat{\mathbf{B}}\left(\mathbf{x_1},0\right)\hat{\mathbf{B}}\left(\mathbf{x_2},0\right)...\hat{\mathbf{B}}\left(\mathbf{x_i},0\right)...\hat{\mathbf{B}}\left(\mathbf{x_N},0\right)\mid\psi\left(t\right)\rangle\end{aligned}
\end{equation}

A straightforward calculation yields
\begin{equation}
  \bm{\psi}^B\left(\mathbf{x},t\right) = \mathrm{i}c\sum_{\lambda=\pm}\int{\mathrm{d}^3k\over (2\pi)^3}(1/\sqrt{\left|\left|\mathbf{k}\right|\right|})\left(\boldsymbol{k}\times\boldsymbol{\epsilon}_{\left(\lambda\right)}\left(\boldsymbol{k}\right)\right)\mathrm{e}^{\mathrm{i}\left(\mathbf{k}\cdot\mathbf{x}-c\left|\left|\mathbf{k}\right|\right|t\right)}c_{\lambda} \left(\mathbf{k},t\right) .
\label{eq:EBP}
  \end{equation}

 It is readily proven   \cite{Goessens} that these single-photon wave functions obey Maxwell's equations:
\begin{subequations}
\begin{eqnarray}
\boldsymbol{\nabla} \times   \bm{\psi}^E(\boldsymbol{x},t) & = & -\partial_t  \bm{\psi}^B(\boldsymbol{x},t) \label{eq:M1} \\
\boldsymbol{\nabla} \times   \bm{\psi}^B(\boldsymbol{x},t) & = &\frac{1}{c^2}   \partial_t \bm{\psi}^E(\boldsymbol{x},t) \label{eq:M2} \\
 \boldsymbol{\nabla} \cdot \bm{\psi}^E(\boldsymbol{x},t) & = & 0 \label{eq:M3}\\
  \boldsymbol{\nabla} \cdot \bm{\psi}^B\boldsymbol{x},t) & = & 0 \label{eq:M4}
\end{eqnarray}
\end{subequations}

It other words, in the vacuum, the single-photon wave function propagates according to Maxwell's equations. This could suggest that Maxwell fields are already quantized to begin with, which is a view shared by certain physicists, and offers a simple and satisfactory picture of single photon optics\footnote{ As was noted in reference \cite{KiesslingZadeh}, {\it...In the quantum optics literature... one often finds the claim that the electromagnetic Maxwell field is the photon wave function, and Maxwell’s field equations are the photon wave equation — in disguise!...} }. However we do not share this interpretation which, at best, should be taken with a grain of salt. As we shall now show, when more than one photon is present, e.g. when photon pairs are produced as is the case in the device presented in figure 2, Maxwell physics does not always suffice to render account of the observations (section \ref{sec2tris}).

\subsection{Maxwell versus Schroedinger: $N$ photon case.\label{Nphot}}As we shall argue here, it is only in particular situations that the quantum wave function ``lives'' in the three dimensional, physical, space. Most often, whenever photons are entangled, a three dimensional representation does not hold anymore and must be replaced by a description in the configuration space.

There exist of course situations where no entanglement is present, and this occurs for instance when light is prepared at time $t=0$ in the monomode Fock state  $\mid\!N\rangle$=${ (\hat{\tilde a}^\dagger)^N\over \sqrt{ N!}}\mid\!N=0\rangle$ (with \cite{SM} an arbitrary ``tilded'' mode $\hat{\tilde a}^\dagger=\sum_{\lambda=\pm}\int\mathrm{d}^3k\,\tilde c_{\lambda}\left(\mathbf{k}\right)\hat{a}_{\left(\lambda\right)}^\dagger\left(\mathbf{k}\right)$).

Then, boson statistical effects imply that the $N$-photon wave electric function factorizes into the product of identical single-photon wave functions:

\begin{equation} \label{NWaveFact}
\bm{\psi}^E\left(\mathbf{x_1},\mathbf{x_2},...,\mathbf{x_N},t\right)=\bm{\psi}^E\left(\mathbf{x_1},t\right)  \cdot  \bm{\psi}^E\left(\mathbf{x_2},t\right)  ... \bm{\psi}^E\left(\mathbf{x_N},t\right)\end{equation}

where each individual (single photon) wave function obeys Maxwell equations. 

Let us for instance show this result in the two photons case ($N$=2). 

Then, (introducing the subscripts $i,j$, with $i,j \in \{x,y,z\}$ indicating that we consider the $i$th Cartesian component of the polarisation in $\mathbf{x_1}$ and the $j$th one in $\mathbf{x_2}$,), the 2-photon wave function reads


\begin{align} \label{2WaveFact}
&(\bm{\psi}^E)^{ij}\left(\mathbf{x_1},\mathbf{x_2},t\right)\equiv\langle 0\!\mid{\hat{\mathbf{E}^i}\left(\mathbf{x_1},0\right)\hat{\mathbf{E}^j}\left(\mathbf{x_2},0\right)\over \sqrt{ 2!}} {(\sum_{\lambda=\pm}\int\mathrm{d}^3k\,\tilde c_{\lambda}\left(\mathbf{k}\right)\hat{a}_{\left(\lambda\right)}^\dagger\left(\mathbf{k},t=0\right)\mathrm{e}^{-\mathrm{i}c\left|\left|\mathbf{k}\right|\right|t})^2\over \sqrt{ 2!}}\mid\!0\rangle\nonumber =\\&{1\over 2}\langle 0\!\mid\mathrm{i}c\cdot(\sum_{\lambda_1=\pm}\int {\mathrm{d}^3k_1\over (2\pi)^3} \sqrt{\left|\left|\mathbf{k_1}\right|\right|}\hat{a}_{\left(\lambda_1\right)}\left(\mathbf{k_1}\right)\bm{\epsilon}^i_{\left(\lambda_1\right)}\left(\mathbf{k_1}\right)\mathrm{e}^{\mathrm{i}\mathbf{k_1}\cdot\mathbf{x_1}})\cdot\mathrm{i}c\cdot(\sum_{\lambda_2=\pm}\int {\mathrm{d}^3k_2\over (2\pi)^3} \sqrt{\left|\left|\mathbf{k_2}\right|\right|}\hat{a}_{\left(\lambda_2\right)}\left(\mathbf{k_2}\right)\bm{\epsilon}^j_{\left(\lambda_2\right)}\left(\mathbf{k_2}\right)\mathrm{e}^{\mathrm{i}\mathbf{k_2}\cdot\mathbf{x_2}})\nonumber\\&(\sum_{\lambda_3=\pm}\int\mathrm{d}^3k_3\,\tilde c_{\lambda_3}\left(\mathbf{k_3},t=0\right)\hat{a}_{\left(\lambda_3\right)}^\dagger\left(\mathbf{k}_3\right)\mathrm{e}^{-\mathrm{i}c\left|\left|\mathbf{k}_3\right|\right|t})(\sum_{\lambda_4=\pm}\int\mathrm{d}^3k_4\,\tilde c_{\lambda_4}\left(\mathbf{k}_4,t=0\right)\hat{a}_{\left(\lambda_4\right)}^\dagger\left(\mathbf{k}_4\right)\mathrm{e}^{-\mathrm{i}c\left|\left|\mathbf{k}_4\right|\right|t})\mid\!0\rangle  \end{align}

Making use of \begin{align} &\langle 0\!\mid \hat{a}_{\left(\lambda_1\right)}\left(\mathbf{k_1}\right)\hat{a}_{\left(\lambda_2\right)}\left(\mathbf{k_2}\right)\hat{a}^\dagger_{\left(\lambda_3\right)}\left(\mathbf{k_3}\right)\hat{a}^\dagger_{\left(\lambda_4\right)}\left(\mathbf{k_4}\right)\mid\!0\rangle=\\\nonumber &\delta^{Kronecker}(\lambda_1-\lambda_3)\delta^{Dirac}\left(\mathbf{k_1}-\mathbf{k_3}\right) \delta^{Kronecker}(\lambda_2-\lambda_4)\delta^{Dirac}\left(\mathbf{k_2}-\mathbf{k_4}\right) \nonumber\\ &+\delta^{Kronecker}(\lambda_1-\lambda_4)\delta^{Dirac}\left(\mathbf{k_1}-\mathbf{k_4}\right) \delta^{Kronecker}(\lambda_2-\lambda_3)\delta^{Dirac}\left(\mathbf{k_2}-\mathbf{k_3}\right),\nonumber \end{align}

we finally get

\begin{align} \label{3WaveFact}&(\bm{\psi}^E)^{ij}\left(\mathbf{x_1},\mathbf{x_2},t\right)=(\mathrm{i}c)^2\cdot\nonumber \\&(\sum_{\lambda=\pm}\int{\mathrm{d}^3k\over (2\pi)^3} \sqrt{\left|\left|\mathbf{k}\right|\right|}\mathrm{e}^{\mathrm{i}\left(\mathbf{k}\cdot\mathbf{x}_1-c\left|\left|\mathbf{k}\right|\right|t\right)}c_{\lambda}\left(\mathbf{k},t=0\right)\bm{\epsilon}^i_{\left(\lambda\right)}\left(\mathbf{k}\right))
(\sum_{\tilde\lambda=\pm}\int{\mathrm{d}^3\tilde k\over (2\pi)^3} \sqrt{\left|\left|\mathbf{\tilde k}\right|\right|}\mathrm{e}^{\mathrm{i}\left(\mathbf{\tilde k}\cdot\mathbf{x}_2-c\left|\left|\mathbf{\tilde k}\right|\right|t\right)}c_{\lambda}\left(\mathbf{\tilde k},t=0\right)\bm{\tilde\epsilon}^j_{\left(\tilde\lambda\right)}\left(\mathbf{\tilde k}\right))
 \nonumber \\&=(\bm{\psi}^E\left(\mathbf{x_1},t\right) )^i \cdot  (\bm{\psi}^E\left(\mathbf{x_2},t\right))^j \end{align}

This explains among others why the statistical distribution of spots (flashes) recorded by Taylor in 1909 \cite{Taylor} with a dimmed light source in a two-slit experiment {\it \`a la Young} reproduces the continuous distribution derived from Maxwell's theory. This is so because the state of thermal light is a mixture of Fock states.

In the case of coherent states, even when light is sent through a beamsplitter, factorisation is still preserved in the sense that, if at the input we inject the coherent state$|\alpha>_{coh.}\equiv e^{-|\alpha|^2/2}\sum_{N=0}^\infty {\alpha^N\over \sqrt{N!}}|N>$ (with $\alpha$ is a complex number), then at the output of the beamsplitter the state of light is a product of a coherent state in the output arm associated to transmission with a coherent state in the output arm associated to reflection (see appendix 2 for more details): $|\Psi>^{out}=|\alpha^{in}\cdot t>_{coh.}^{Transmitted}\cdot |\alpha^{in}\cdot r>_{coh.}^{Reflected}$, where $t$ and $r$ represent the (complex) amplitudes of transmission and reflection. Here again the $N$ photons electric and magnetic wave functions associated to a Fock state $|N>$ are the products of $N$ identical single photon (3D) wave functions, which are solutions of Maxwell's equations as we have shown before.

All this also explains why in many situations (where thermal and laser light sources are used for instance) Maxwell's theory suffices to render account of the observations. 

However, whenever we consider at least two photons, entanglement is likely to be present, and a classical description {\it \`a la} Maxwell in the physical, 3-d space must be replaced by a description in the configuration space. This is for instance the case in the HOM experiment which cannot be explained in terms of Maxwell fields as we now show\footnote{A wave function approach to describe photons is not common. In the majority of standard text-books the HOM experiment is formulated in terms of creation-destruction operators associated to plane wave modes. Such an approach emphasizes the corpuscular nature of light. We consider however that a wave function approach is more precise because plane waves do not exist in nature. It also emphasizes the fact that we are dealing with processes occuring in space-time or configuration space-time, which is generally lacking in approaches based on Fock states\cite{Gordon}.}.
\subsection{HOM experiment.\label{sec2tris}}
Let us assume here that a well-chosen source, e.g. a pumped non-linear crystal, makes it possible to prepare pairs  of equivalent photons, the so-called idler and signal photons. What we mean hereby is that 

(i) the shape of the wave function associated to each photon is the same and 

(ii) they are sent at the same time. 

Let us denote $\Psi^{idler}(x,t)$ and $\Psi^{signal}(y,t)$ these wave functions (here we consider that light is always polarized orthogonally to the plane containing the photon trajectories; the single photon electric field/wave function is thus essentially a complex scalar function). The choice of the variables $x,y$ is dictated by the fact that the idler (resp. signal) photon propagates horizontally (resp. vertically) at the output of the source (see figure \ref{fig2b}). The wave function $\Psi_2^{in}$ of the pair before entering the beamsplitter then reads ${e^{i\delta \theta}\over \sqrt 2}(\Psi^{idler}(x_1,t)\Psi^{signal}(y_2,t)+\Psi^{idler}(x_2,t)\Psi^{signal}(y_1,t))$, where we took account of the bosonic nature of the photons and symmetrised the wave function accordingly. Already at this level, entanglement is present due to bosonic indiscernibility. After passing through the 50-50 beamsplitter, the two photon wave function $\Psi_2^{out}$ at the output of the HOM device represented in figure \ref{fig2b}) is thus (up to an irrelevant global phase that we shall no longer write in what follows)

${1\over 2\sqrt 2}[\{(\Psi^{idler}(x_1,t)+i\Psi^{idler}(y_1,t))(i\Psi^{signal}(x_2,t)+\Psi^{signal}(y_2,t))\}$

$+\{(\Psi^{idler}(x_2,t)+i\Psi^{idler}(y_2,t))(i\Psi^{signal}(x_1,t)+\Psi^{signal}(y_1,t))\}]$

Now, $\Psi^{signal}(x,t)=\Psi^{idler}(x,t)=\Psi(x,t)$ and $\Psi^{signal}(y,t)=\Psi^{idler}(y,t)=\Psi(y,t)$ because the source produces identical photons so that finally

$\Psi_2^{out}={i\over \sqrt 2}\{\Psi(x_1,t)\Psi(x_2,t)+\Psi(y_2,t)\Psi(y_1,t)\}$.

This is the essence of HOM experiment \cite{HOM}: due to the indistinguishability and equivalence of the two photons, they always leave the beamsplitter through the same output port (horizontal xor vertical). In other words, either there are two photons in the vertical output arm xor there are two photons in the horizontal one. This occurs of course provided they are perfectly synchronised.  The probability of joint firing of the detectors 1 and 2 thus decreases when the delay between the two incoming photon goes to zero (this is the so-called HOM dip). This coalescence phenomenon is called photon bunching. It is an example of interference in configuration space. The left part of the device of figure \ref{fig2a} actually constitutes a HOM device (fig. \ref{fig2b}) provided the source prepares equivalent photons. 

Note that the state $\Psi_2^{out}$ is maximally entangled, as well as the state $\Psi_2^{in}$. The two photons produced by the crystal are identical but their supports do not overlap, and their entanglement results from their undistinguishability (see also Ref. \cite{Shih} for a similar analysis in terms of biphotons). 

\subsection{Photon trajectories.}

The local quantum Poynting vector reads, in the single photon case,  $\frac{1}{2\mu_0}((\bm{\psi}^E\left(\mathbf{x},t\right))^*\times\bm{\psi}^B\left(\mathbf{x},t\right)+\bm{\psi}^E\left(\mathbf{x},t\right)\times (\bm{\psi}^B\left(\mathbf{x},t\right))^*)$.

 In full analogy with classical Maxwell's equations, one can also show that the density $\frac{\epsilon_0}{2}\left[\vert \bm{\psi}^E\left(\mathbf{x},t\right)\vert^2+c^2\vert \bm{\psi}^B\left(\mathbf{x},t\right)\vert^2\right]$ of the photonic population at a given place $(x,y,z)$ and a given time $t$, obeys the conservation equation

 \begin{equation}
 \frac{\partial }{\partial t}(\frac{\epsilon_0}{2}\left[\vert \bm{\psi}^E\left(\mathbf{x},t\right)\vert^2+c^2\vert \bm{\psi}^B\left(\mathbf{x},t\right)\vert^2\right])+div.\frac{1}{2\mu_0}((\bm{\psi}^E\left(\mathbf{x},t\right))^*\times\bm{\psi}^B\left(\mathbf{x},t\right)+\bm{\psi}^E\left(\mathbf{x},t\right)\times (\bm{\psi}^B\left(\mathbf{x},t\right))^*)= 0
 \end{equation}

In analogy with classical fluid dynamics, it is natural to define a local velocity as follows $\bm{v}\left(\mathbf{x},t\right)$:

$\bm{v}\left(\mathbf{x},t\right)={\frac{1}{2\mu_0}((\bm{\psi}^E\left(\mathbf{x},t\right))^*\times\bm{\psi}^B\left(\mathbf{x},t\right)+\bm{\psi}^E\left(\mathbf{x},t\right)\times (\bm{\psi}^B\left(\mathbf{x},t\right))^*)
\over (\frac{\epsilon_0}{2}\left[\vert \bm{\psi}^E\left(\mathbf{x},t\right)\vert^2+c^2\vert \bm{\psi}^B\left(\mathbf{x},t\right)\vert^2\right])}$, which may be interpreted as the single photon guidance equation. The generalization to  more than one photon is straightforward.

 \section{Generation and detection of empty waves revisited: pilot wave dynamics in  configuration space and effective collapse.\label{sec3}}
Let us now reconsider the experimental proposal outlined in section \ref{sec1} (fig. \ref{fig2a}).  As we mentioned by then, if we make abstraction of the presence of detectors 3 and 4, the first part of the set up (in the green frame) is equivalent to a Hong Ou Mandel (HOM) device (fig. \ref{fig2b}).  Then, as we have shown in section \ref{sec2tris}, whichever value $\delta \theta$ could take, it is predicted in a NCI or in a 6D CI {\it \`a la} Bohm that the two equivalent photons (idler and signal photons) produced at the level of the crystal always appear in pair at the output of the HOM device and the two photon wave function, just before passing through the output beamsplitter of the MZ device (fig. \ref{fig2b}), in the blue frame of fig. \ref{fig2a}, is equal (up to an irrelevant global phase $e^{i\delta \theta}$ that we will no longer write here) to ${i\over \sqrt 2}\{\Psi(x_1,t)\Psi(x_2,t)+e^{i\delta \phi}\Psi(y_2,t)\Psi(y_1,t)\}$.


Therefore, when the dephasing $\delta \phi$ between both arms inside the MZ is equal to 0, the wave function $\tilde \Psi_2^{out}$ at the output of the MZ device, before arriving to detectors 1 and 2 obeys

\begin{align}&\tilde \Psi_2^{out}={i\over 2\sqrt 2}\{(\Psi(x_1,t)+i\Psi(y_1,t))(i\Psi(y_2,t)+\Psi(x_2,t))+(\Psi(y_2,t)+i\Psi(x_2,t))(i\Psi(x_1,t)+\Psi(y_1,t))\}\nonumber\\&
={(-1)\over \sqrt 2}\{\Psi(x_1,t)\Psi(y_2,t)+\Psi(x_2,t)\Psi(y_1,t)\},\end{align} 
where $x$ ($y$) is here assigned to the horizontal (vertical) output arm leading to detector 2 (1).

This means that, contrary to the HOM device (bunching), the two photons never leave the MZ device through the same output port (anti-bunching)\footnote{Actually when $\delta \phi$  varies from 0 to $\pi$ an interference pattern of visibility 100 percent is predicted to occur, $\delta \phi=\pi$ corresponding to bunching as in the HOM experiment. This corroborates the observations reported in Ref. \cite{Bristol} where, due to some experimental unperfections such as misalignments, desynchronisations and so on a visibility of 72 percent has been achieved for a similar set up. In this case, remarkably, the equivalent photons were obtained from two different sources, in a regime of pulsed excitation.}. 

Before pursuing our analysis, let us evaluate the full wave function in presence of detectors 3 and 4, gathering the computational bricks previously developed in this paper. With probability 1/4, a pair of photons will enter the MZ device, with probability 1/4, only the signal photon will do so,  with probability 1/4, only the idler photon will do so, and finally, with probability 1/4, one photon will be present in the channel of detector 3 and the other one in the channel of detector 4, in which case no photon enters the MZ device. For perfect detectors a click will certainly occur when a photon is present in the channel of a detector but this does not really matter, our analysis being still valid in the case of unperfect detectors.

The corresponding wave function $\tilde{\tilde \Psi}_2^{out}$, at the output of the device of figure \ref{fig2a} thus obeys

\begin{eqnarray}\tilde{\tilde \Psi}_2^{out}={(-1)\over 2\sqrt 2}\{\Psi(x_1,t)\Psi(y_2,t)+\Psi(x_2,t)\Psi(y_1,t)\nonumber \\
+\Psi(x_1,t)\Psi(u_2,t)+\Psi(x_2,t)\Psi(u_1,t)+\Psi(y_1,t)\Psi(v_2,t)+\Psi(y_2,t)\Psi(v_1,t)\nonumber \\
-\Psi(u_1,t)\Psi(v_2,t)-\Psi(u_2,t)\Psi(v_1,t)\},\label{full}\end{eqnarray} 

where the first line corresponds to both photons entering the MZ device, the second line to a single photon doing so, and the last line to no photon inside the MZ device. Here again, $x$ ($y$) is assigned to the horizontal (vertical) output arm leading to detector 2 (1), while $v$ and $u$ are assigned to detectors 3 and 4 respectively.

If only one photon enters the MZ device, which occurs when the other photon entered the channel of detector 3 (xor detector 4), the 3D single photon description performed in section 1 implies that the MZ device will be transparent, when there is a collapse in detector 3 (xor 4) but also when there is no collapse, provided we apply de Broglie-Bohm interpretation to trajectories in configuration space. The reason why this also happens in absence of collapse is that, as is well-known in the case of de Broglie-Bohm theory, there is a click in a detector only if a photon is present in the corresponding channel. Moreover, when a photon is present in the channel of this detector  (3 xor 4), the guidance equation of the other photon is the same as the guidance equation of the collapsed state. A click in detector 3 (or 4) thus results into an ``effective'' collapse at the level of the pilot wave assigned to the other photon.

Actually a very similar effective collapse is also predicted to occur if we consider the Bohmian Conditional Wave Function associated to the sub-system of a larger system e.g. the entire universe \cite{NorsenUni}. Here the subsystem is a pair of photons and the larger system contains the detectors.

To conclude this section, it is worth noting that

1) if a click in detector 3 xor 4 occurs, the statistical distribution assigned to clicks in detectors 1 and 2 is the same in presence or in absence of collapse (provided we adopt the in the latter case the NCI {\it \`a la} Bohm based on the guidance equation in the 6D configuration space\footnote{We expect this result to hold in all possible experimental configurations: NCI {\it \`a la} Bohm will always lead to the same predictions as standard CI. From this point of view, they are ad hoc, a well-known feature of de Broglie-Bohm interpretation.}).

2) this statistical distribution differs from the distribution associated to the NCI  {\it \`a la} de Broglie outlined in the section  \ref{sec1}, where the pilot wave was treated as a 3D object. 

Henceforth, the device presented in figure \ref{fig2a} makes it possible to discriminate both approaches experimentally. It constitutes from this point of view a crucial test of the empty wave approach proposed by Croca and coworkers \cite{Croca,Crocabook}.

As we already mentioned in the introduction, the NCI based on the guidance equation in configuration space is in the line of Bohm's views about ``hidden'' quantum trajectories. The NCI  outlined in the section 1, where the pilot wave is treated as a 3D object fits, according to us, to de Broglie's considerations about the pilot wave interpretation, which suggests a fine structure in the pilot wave interpretations, between the 3D NCI {\it \`a la} de Broglie and the 6D NCI {\it \`a la} Bohm\footnote{This distinction was explicitly recognized by F.Selleri who wrote, about de Broglie's picture, more than 30 years ago  \cite{Selleri}, the following: {\it...According to this picture a quantum object is composed of a small particle which is constantly localized in space, and of an objective real wave $\phi(x,y,z,t)$ which is a physical process propagating in space and time...}In the same paper, F.Selleri wrote, about Einstein's photon that {\it...A problem immediately coming to mind with Einstein's philosophy is the following: If the localized particle carries all the energy and momentum, in which sense can the wave be considered real? This problem was felt so acutely by Einstein that he referred to these waves a Gespenterfelder (ghost fields): An object without energy and momentum is in fact unable to exert a pressure when impinging on a material surface, which means that it does not have that quality that makes us call something real. Still, the equations of the quantum theory describe this wave as propagating in space and time. The difficulties associated with the concept of an {\bf empty wave}(...) have led many people to discard the idea as a scientific impossibility(...). It will be shown in the present chapter that the previous objection can be overcome because not only changes in energy and momentum can be observed, but modifications of probabilities as well.} The last sentence also applies to the present paper.}as we explain now more in detail, completing the analysis already carried out in Ref. \cite{Intro}.

\section{Bohm versus de Broglie:  a proposed fine structure in pilot wave dynamics.\label{sec4}}
In 1926 \cite{1926}, de Broglie laid the foundations of an interpretation of quantum mechanics
known today  as the de Broglie-Bohm interpretation. One of the basic principles of this interpretation
is that every quantum system is characterized by a continuous spatio-temporal trajectory. In this approach, the Heisenberg uncertainty relations
are not considered as an intrinsic limitation to the concept of trajectories but rather as
a constraint relative to the statistical distributions of the position and
the momentum of a quantum system. 
In the 50's Bohm will tackle the measurement problem (still pretty much disregarded
in 1926) and suggest a solution anticipating the
concept of decoherence, developed later in the 1970s within the framework of open quantum systems \cite{Wheeler}. The
decoherence idea completes the theoretical construction prefigured by
Broglie in 1926 and 1927 (during his presentation at the 1927 Solvay
Conference \cite{1927} de Broglie gave a simplified version of his Double-Solution program \cite{CDWannales}, that constitutes the backbone of what is known
today as the ``pilot wave theory'' \cite{Holland,Bricmont}) along the following lines: 
 
 -the particles are localized at all times in a spatial region  much smaller  than the support of the Schr\"odinger  wave;
 
 - the trajectories satisfy the {\bf guidance condition} applied to the 
 Schr\"odinger pilot wave ($\Psi$ wave) (here expressed in the $N$ particles case) according to which the 3N dimensional velocity obeys
  \begin{align}&\overrightarrow{v}=\label{g}\\\nonumber &{\hbar\over m}{Im.(\Psi^*(x_1,y_1,z_1,x_2,y_2,z_2,...,x_i,y_i,z_i,...,x_N,y_N,z_N, t)\overrightarrow{\nabla} \Psi(x_1,y_1,z_1,x_2,y_2,z_2,...,x_i,y_i,z_i,...,x_N,y_N,z_N, t)\over |\Psi(x_1,y_1,z_1,x_2,y_2,z_2,...,x_i,y_i,z_i,...,x_N,y_N,z_N, t)|^2}\end{align}
 
 
 -at all times the statistical distribution of positions is governed by
 Born's rule (that is $|\Psi(x_1,y_1,z_1,x_2,y_2,z_2,...,x_i,y_i,z_i,...,x_N,y_N,z_N, t)|^2$);

 - any measurement is ultimately a measurement of position. This means among others that a detector will click if and only if the particle ``is located inside'' and/or ``passes through'' the detector.
 
Already in his seminal paper of 1926 \cite{1926}, de Broglie faced a serious problem: when one
considers a composite quantum system (for example a pair of
particles) described by a non-factorizable Schr\"odinger wave function
(``entangled'', according to the term introduced by Schr\"odinger  in 1935), the
interpretation leads to trajectories that, in agreement with equation (\ref{g}), are not defined in our 3-dimensional physical space, but in configuration space (having
dimension 3N for a system composed of N particles). de Broglie
will never really overcome this problem and neither will he  subscribe \cite{dBjalon} to the view developed later by Bohm (in 1951) \cite{bohm521,bohm522}  and
highlighted by Bell (in 1964) \cite{bell}, a view according to which quantum physics is an inherently non-local theory. In the 1950s he will also strive to develop \cite{CDWannales} with Andrade e Silva \cite{acasci1,acasci2,acasci3,acasci4}, a version of the pilot wave interpretation in
which the interactions between quantum systems are defined not
in configuration space but in $R^3$ (see also Ref.\cite{Norsen2} for a more recent approach).

When Bohm updated the pilot wave interpretation in 1952 \cite{bohm521,bohm522}, he was the first to reformulate the EPR paradox in terms of spin 1/2 particles, a result which played an essential role in the genesis of Bell's inequalities as Bell recognized later. From this point of view, Bohm was the first (and not de Broglie) to appreciate and to fully recognize the non-local character of quantum correlations between entangled systems, which is manifest at the level of the guidance equation (\ref{g}). He also understood in depth how one could get rid of the very polemical projection postulate in the framework of the pilot wave interpretation. This is so indeed because when a microscopic quantum system interacts with a macroscopic measurement apparatus, they will get strongly entangled, so that the reduced density matrix of the microscopic quantum system will become a decoherent mixture. This ingredient explains why, FAPP, empty waves will never give rise to interferences between the branch of the wave function associated to the click in the detector and other branches. It explains thus why it is impossible to observe empty waves in simple situations where the microscopic quantum system under study is disentangled, before the measurement process, from the rest of the world. If we wish to consider the more general situation during which the system under study is entangled with another system (like in the present paper where we consider pairs of entangled photons), the decoherence ingredient of Bohm, taken together with the aforementioned effective collapse mechanism (section \ref{sec3}), makes it possible to formulate the pilot wave interpretation without invoking at all the projection postulate. It is for all these reasons that we attached the label 3$N$D NCI to Bohm's formulation of the pilot wave interpretation  and the label 3D NCI to the formulation of de Broglie and his followers. It is worth noting to conclude this more historical and bibliographical section that the paradigmatic version of 3D NCI outlined in section 1 (see also section \ref{sec2bis}), where photons follow the Poynting-Maxwell lines of flow, is widely spread among classical physicists\footnote{One can read for instance in the standard text-book of Sakurai \cite{Sakurai} { \it...The classical limit of the quantum theory of radiation is achieved when the number of photons becomes so large that the occupation number may as well be regarded as a continuous variable. The space-time development of the classical electromagnetic wave approximates the dynamical behavior of trillions of photons....}}. The density of photons would be equal in this approach to the Poynting energy density divided by the individual energy of a single photon, $h\nu$. The empty wave model of section 1 perfectly fits into this scheme. The description of the trajectories associated to the passage through a barrer, very similar to those in a beamsplitter, has for instance been studied  in the pilot wave approach, by Dewdney and Hiley \cite{Dewdney,Holland}. The de Broglie-Bohm trajectories in a Mach-Zehnder device also received some attention in connection with ``surrealistic trajectories''  \cite{Vaidman} and with Wheeler's paradox \cite{Hiley} but it is worth noting that all these works remained limited to the single photon case in which case empty waves are defined over the physical 3-D space and not over a larger configuration space, in which case we expect that the predictions made in the framework of the de Broglie-Bohm interpretations are equivalent to those made in the framework of the Copenhagen interpretation \cite{Hardy}. In many works about de Broglie-Bohm trajectories it is tacitly assumed from the beginning that the predictions of the de Broglie-Bohm interpretations are always equivalent to those made in the Copenhagen interpretation (in other words the dB-B interpretation is ad hoc). As we have shown here, this seems to be also the case when more than one particle is present, provided the dB-B dynamics is defined over the configuration space, even in presence of detectors (see also Ref.\cite{NorsenUni}), due to the aforementioned effective collapse mechanism (section \ref{sec3}).
\section{Conclusions.\label{sec5}}
Roughly summarized the main conclusions of our analysis are

1) in configuration space, the influence of empty waves vanishes, and it is impossible to discriminate between CI and 3$N$D NCI {\it \`a la} Bohm. In the last resort, this is so because in 3$N$D NCI {\it \`a la} Bohm, an {\it effective} collapse process is present which is equivalent FAPP to the projective collapse {\it \`a la} von Neumann.

2) the validity of 3D NCI {\it \`a la} de Broglie outlined in section \ref{sec1}, and its corollary, the existence of empty waves, would in turn establish the validity of a hybrid model, neither classical nor quantum, where the pilot wave is a 3D object. Associating trajectories of photons to lines of flow of classical, 3D, Maxwell fields  constitutes a paradigmatic example of this class of hybrid models.

Henceforth, the existence of empty waves of the type described here would therefore NOT establish (the validity of) de Broglie-Bohm theory stricto sensu. It would rather falsify the quantum theory as a whole and indicate that it must be replaced by a hybrid interpretation where particles are guided by 3D fields. These experiments remain crucial in our eyes, but they do not make it possible to discriminate between CI and NCI. They rather make it possible either to falsify hybrid models (that we baptised under the label 3D NCI {\it \`a la} de Broglie\footnote{Note that, following F.Selleri \cite{Selleri}, we could as well have baptised these waves Einstein-de Broglie waves in order to reinforce the connection with some ideas of the young Einstein about the photon. Einstein-de Broglie waves live in 3 +1 dimensions, while de Broglie-Bohm live in 3N+1 dimensions}) where 3D fields (4D if we include time to the description of the fields) only are present, as e.g. in Maxwell's theory and in general relativity theory, or to falsify the quantum theory as a whole (including ``strict'' 3$N$D NCI {\it \`a la} Bohm and CI). 

In a certain sense, the HOM experiment \cite{HOM} already emphasised in the past the non-classical nature of quantum correlations, and the proposals in section \ref{sec1} and appendix 1 are to some extent reminiscent of HOM's analysis. The present paper has the merit however to shed a new light onto the prior role played by the configuration space in pilot wave dynamics. It is not the first time that the debate is brought to the realm of experimentalists actually: Mandel and coworkers already realized an experiment aimed at detecting empty waves in the past, 30 years ago \cite{Mandel2}. This experiment has been criticised however  \cite{Crocacomment} and it was not fully convincing. In particular, it was conceived in such a way that if empty waves existed, it would have led to the appearance of interferences. Moreover, it has been claimed that partial interference was observed anyhow \cite{Crocabook}, which justifies in our eyes to propose to realize new experiments. This is why we devoted time and energy in the present paper to discuss these questions again. Even if NCI {\it \`a la} de Broglie do not explain integrally the observations, and that the influence of empty waves is only partial, their manifestation would have far reaching consequences for what concerns our interpretation of the quantum theory. Moreover, it could present appealing technological implications. Indeed, as suggested in Ref.\cite{Croca}, if empty waves are used to carry information, they are likely to suffer less from dissipation than their non-empty, corpuscular counterpart. A potential application would be long distance quantum communication, which justifies the theoretical bet made here concerning the possible existence of an ``empty wave effect''.

\begin{figure}[hhh!]\centering\includegraphics[scale=0.35]{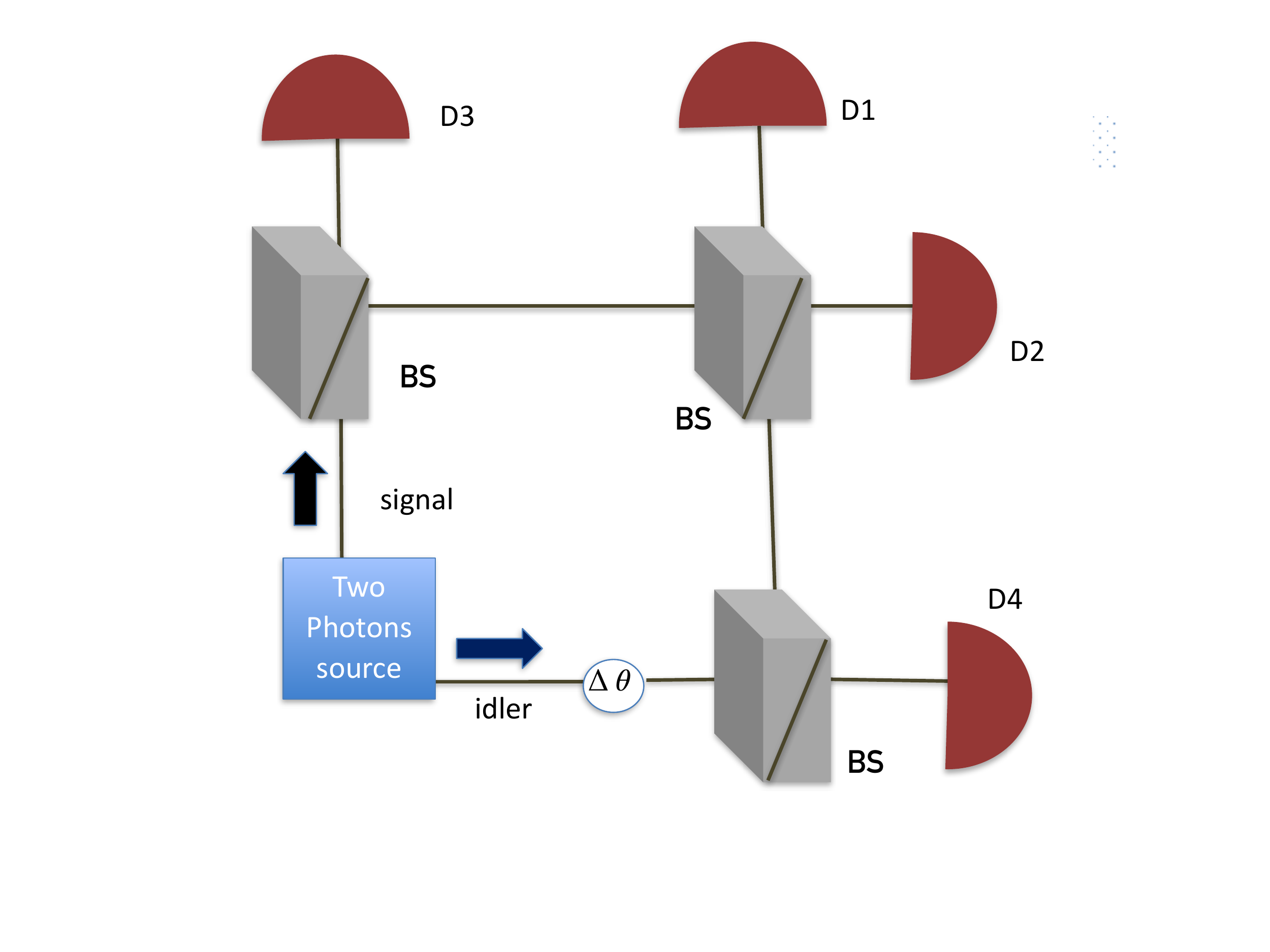}\caption{Simplified scheme for detecting empty waves.}\label{figappA}\end{figure}

\section*{Appendix 1: testing phase coherence between the signal and idler waves experimentally.}Let us consider the scheme presented in figure \ref{figappA}. It consists in suppressing in the full device in figure \ref{fig2a} the beamsplitter at the output of the MZ device. In other words, we should just keep the HOM device in the green  frame of figure \ref{fig2a} plus detectors 3 and 4 aimed at generating empty waves. If no 3D empty wave is present, the value of $\delta \theta$ is irrelevant because it plays the role of a global phase which is like nothing in the standard quantum theory as is well-known. This is obvious if we note that  the wave function of the pair of photons, just before entering the beamsplitters connecting to the detectors 3 and 4 then reads ${1\over \sqrt 2}e^{i\delta \theta}(\Psi^{idler}(x_1,t)\Psi^{signal}(y_2,t)+\Psi^{idler}(x_2,t)\Psi^{signal}(y_1,t))$. There is thus no way to estimate the value of $\delta \theta$ in NCI, and the same results can be shown to hold in 6D CI {\it \`a la} Bohm. If there is a click in detector 3 xor 4, the probability that detector 1 fires is equal to the probability that detector 2 fires and both are equal to 1/4, whichever the value of $\delta \theta$ could be. We predict the same behaviour if we describe this situation in the framework of 3D CI {\it \`a la} de Broglie in absence of phase coherence between the idler and signal pulses. If some coherence, even partial coherence, is present however, the presence of empty waves results into a partial interference pattern when $\delta \theta$ gets varied from 0 to $2\pi$.


In particular, if $\theta$ is equal to $\pm\pi/2$, in the case of maximal coherence, the surviving photon will always leave the beamsplitter along the same arm, as we noted previously. In such an experiment it is even not necessary to recombine the two channels at the outputs of the first beamsplitter...As was noted by Croca {\it et al.} \cite{Croca} however, the visibility of the interference pattern is zero whenever the two photon pulses are incoherent in phase, in which case we recover the predictions made in the framework of NCI, and the scheme proposed here is useless. It is however worth doing the experiment in order to estimate the degree of coherence between the signal and idler pulses to begin with. If coherence is minimal it is not conclusive but otherwise this would in itself constitute a crucial experiment. In a sense, empty waves would also make it possible then to measure a global phase, which is impossible mission in standard quantum mechanics.

\section*{Appendix 2: replacing the PDC source by a laser source.}As explained in section  \ref{Nphot}, {\it ... in many situations (where thermal and laser light sources are used for instance) Maxwell's theory suffices to render account of the observations... } This emphasises the role played by the optical source in these experiments. If we  send for instance a Fock state $|E_N>$ at the input of a beamsplitter  then at its outputs we get \cite{DD} the state $\sum_{M=0}^N(|t|^2)^{N-M}(|r|^2)^{M}\sqrt{{N !\over M!\cdot (N-M) !}}|E^{transmitted}_{N-M}>|E^{reflected}_{M}>$, where $t$ and $r$ represent the amplitudes of transmission and reflection. Moreover, the $N$-photons wave function associated to a Fock state can be shown to factorize into products of identical one photon wave functions as shown py us in the section \ref{Nphot}.

If a coherent state $|\alpha>_{coh.}\equiv e^{-|\alpha|^2/2}\sum_{N=0}^\infty {\alpha^N\over \sqrt{N!}}|E_N>$ enters the beamsplitter, then the output state obeys

 $|\Psi>^{out}=e^{-|\alpha|^2/2}\sum_{N=0}^\infty {\alpha^N\over \sqrt{N!}}\sum_{M=0}^N(|t|^2)^{N-M}(|r|^2)^{M}\sqrt{{N !\over M!\cdot (N-M) !}}|E^{transmitted}_{N-M}>|E^{reflected}_{M}>$

$=e^{-|\alpha|^2(|t|^2+|r|^2)/2}\sum_{N=0}^\infty {\alpha^M\alpha^{N-M}\over \sqrt{N!}}\sum_{M=0}^N(|t|^2)^{N-M}(|r|^2)^{M}\sqrt{{N !\over M!\cdot (N-M) !}}|E^{transmitted}_{N-M}>|E^{reflected}_{M}>$

$=(e^{-|\alpha|^2|t|^2/2}\sum_{N-M=0}^\infty (\alpha |t|^2)^{N-M}\sqrt{{1\over (N-M) !}}|E^{transmitted}_{M-N}>)(e^{-|\alpha|^2|r|^2/2}\sum^\infty_{M=0}(\alpha |r|^2)^{M}\sqrt{{1\over M!}}|E^{reflected}_{M}>$)

$=|\alpha^{in}\cdot t>_{coh.}^{Transmitted}\cdot |\alpha^{in}\cdot r>_{coh.}^{Reflected}$,

which is is a product of a coherent transmitted state with a coherent reflected state.

\begin{figure}[hhh!]\centering\includegraphics[scale=0.35]{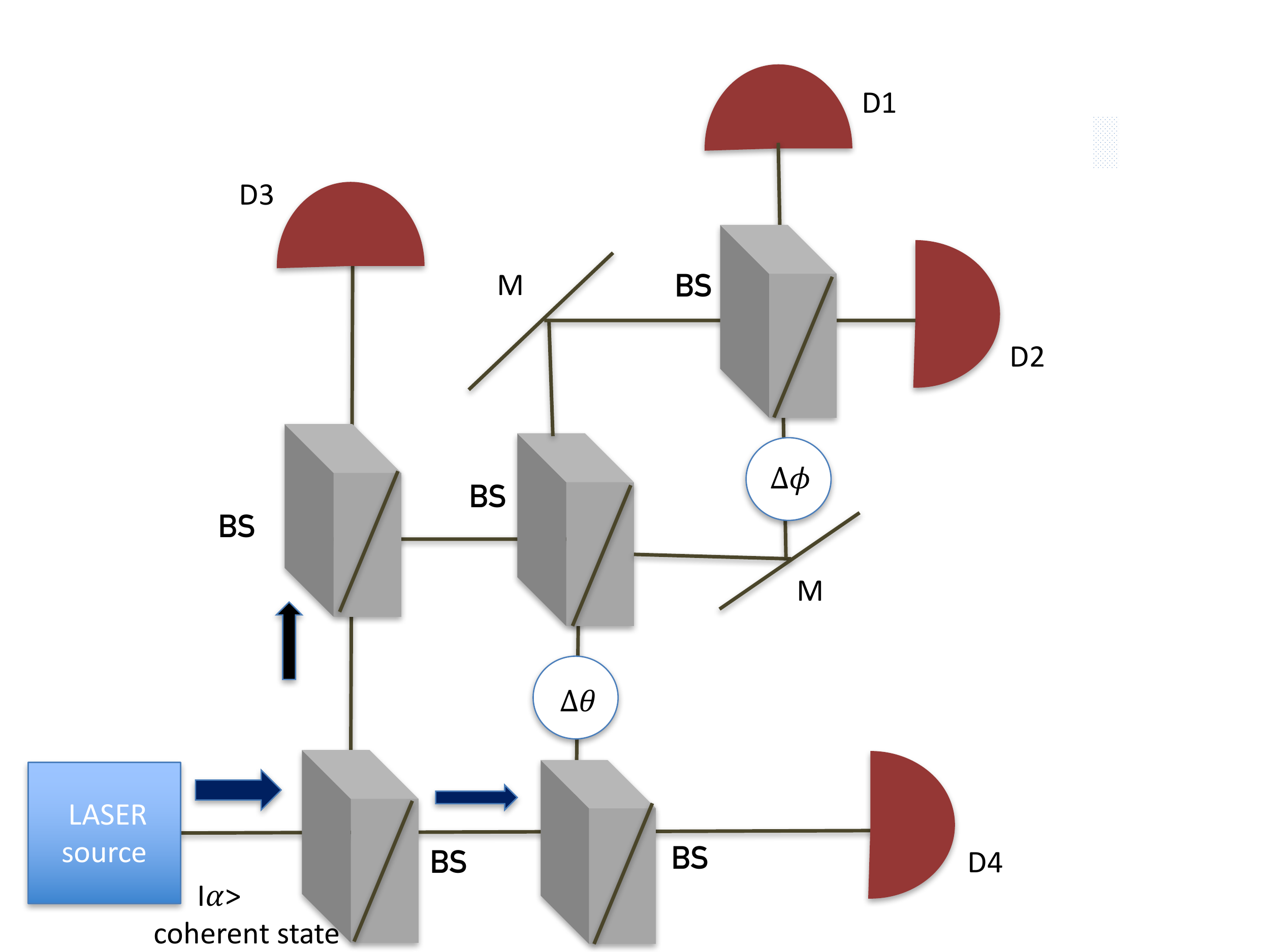}\caption{Replacing the PDC source by a laser source.}\label{figappB}\end{figure}

If we use a laser source coupled to a beamsplitter (as plotted in figure \ref{figappB}) instead of a source of two equivalent photons (e.g. a crystal where a single photon of the pump gets replaced, during a Parametric Down Conversion process, by two equivalent photons, the signal and idler photons), the wave function $\ \Psi^{out}$ thus obeys, instead of equation (\ref{full}), the following equation:

\begin{eqnarray}\Psi^{out}=|\alpha^1>_{coh.}|\alpha^2>_{coh.}|\alpha^3>_{coh.}|\alpha^4>_{coh.}\end{eqnarray} 
where the source produces a coherent state $|\alpha>_{coh.}\equiv e^{-|\alpha|^2/2}\sum_{N=0}^\infty {\alpha^N\over \sqrt{N!}}|N>$, while the complex amplitudes assigned to the coherent states present in outcome channels 1,2,3,4 are equal (up to a global phase) to 

$\alpha_1={i\alpha\over 4}(-1+e^{i\delta \theta}+e^{i\delta \phi}+e^{i(\delta \theta+\delta \phi})$ (detector 1), 

$\alpha_2={i\alpha\over 4}(-1+e^{i\delta \theta}-e^{i\delta \phi}-e^{i(\delta \theta+\delta \phi})$ (detector 2),

$\alpha_3=i\alpha/2$ (detector 3), 

$\alpha_4=\alpha/2$ (detector 4).

This distribution is the same that would be obtained in the framework of Maxwell's theory, and in the framework of NCI {\it \`a la} de Broglie and {\it \`a la} Bohm as well. It is also fully consitent with a CI because the state of light is factorisable (it is a product of coherent states localized at the level of each of the 4 outcome channels/detectors). Therefore the clicks at the levels of detectors 1,2,3,4 are uncorrelated (statistically independent). This explains why no crucial experiment aimed at discriminating between these various interpretational schemes is conceivable when the source is a laser state. These results could appear to be useful however in case they are used to {\bf calibrate} the device described in figure \ref{fig2a}. 

\bibliography{singlephotonfundamentalsv2}

\end{document}